# Cold Atmospheric Plasma Sterilization of FFP3 Face Masks and Long - Term Material Effects

*Alisa Schmidt[1], Chen-Yon Tobias Tschang[1], Joachim Sann[2], Markus H. Thoma[1]*

[1] I. Physical Institute, Justus Liebig University, Heinrich-Buff-Ring 16, 35392 Giessen, Germany

[2] Physical Chemical Institute, Center for Material Research, Justus Liebig University, Heinrich-Buff-Ring 17, 35392 Giessen, Germany

______________________________________________________________________________

## Abstract

The use of cold atmospheric plasmas (CAP) to sterilize sensitive surfaces is an interesting new field of applied plasma physics. Motivated by the shortages of face masks and safety clothing at the beginning of the corona pandemic, we conducted studies on the sterilization of FF3 face masks with CAP and the resulting material effects.

Therefore, the bactericidal and sporicidal efficacy of CAP afterglow sterilization of FFP3 mask material was investigated by inoculating fabric samples with test germs *Escherichia coli* (*E. coli*) and *Bacillus atrophaeus* (*B. atrophaeus*) and subsequent CAP afterglow treatment in a surface-micro-discharge (SMD) plasma device. In addition, a detailed analysis of the changes in long-term plasma treated (15h) mask material and its individual components - ethylene vinyl acetate (EVA) and polypropylene (PP) - was carried out using surface analysis methods such as laser microscopy, contact angle measurements, X-ray photoelectron spectroscopy (XPS) as well as fabric permeability and resistance measurements.

The experiments showed that *E. coli* and *B. atrophaeus* could both be effectively inactivated by plasma treatment in nitrogen mode (12 $kV_{pp}$, 5 kHz). For *B. atrophaeus* inactivation of more than 4-log was achieved after 30 minutes. *E. coli* population could be reduced by 5-log within one minute of CAP treatment and after five minutes a complete inactivation (> 6-log) was achieved. Material analysis showed that long-term (> 5 h) plasma treatment affects the electrostatic properties of the fabric.

From this it can be deduced that the plasma treatment of FFP3 face masks with the CAP afterglow of an SMD device effectively inactivates microorganisms on the fabric. FFP3 masks can be plasma decontaminated and reused multiple times but only to a limited extent, as otherwise the permeability levels no longer meet the DIN EN 149 specifications.

# 1 Introduction

The outbreak of the Covid-19 pandemic at the end of 2019 has posed a great challenge for our society, especially regarding the health care sector, which we are still dealing with one and a half years later. In December 2019, a large number of novel coronavirus disease cases were first identified in Wuhan, Hubei province, China, and from there rapidly spread around the globe.

Scientific research suggests that the severe respiratory syndrome coronavirus (SARS-CoV-2) most likely originated in bats and was transmitted to humans. For example, Cui et al. [1] demonstrated via collected data on genetic evolution, receptor binding and pathogenesis that SARS-CoV most likely originated in bats through sequential recombination of bat SARSr-CoVs. Those infected with coronavirus either remained asymptomatic or showed various non-specific reactions to the virus. By January 2$^{nd}$ 2020 Huang et al. researched a group of 41 clearly identified SARS-CoV-2 infected patients in Wuhan. The group consisted of 73% men and 27% women. 32% had underlying diseases and median age was 49. Common reported symptoms included fever (98%), cough (76%), dyspnoea (55%) and myalgia or fatigue (44%), while less common symptoms included sputum production (28%), headache (8%), haemoptysis (5%) and diarrhea (3%) [2]. The rapid spread of SARS-CoV-2 between humans happened via respiratory droplets, as the main way of transmission [3] [4]. The importance of mask wearing to prevent the spread of aerosol transmissible diseases has been reported multiple times. For example, Milton et al. [5] found by molecular methods that total viral copies (influenza) were 8.8 times more numerous in fine (≤5 μm) than in coarse (>5 μm) aerosol particles, which also included infectious virus, and that surgical masks reduced the total number of detected RNA copies by 3.4-fold. Therefore, besides social-distancing, face mask wearing in combination with proper hand hygiene were identified as effective measures to prevent SARS-CoV-2 transmission. Because of these findings and with China being the major facemask producer in the world, contributing to 50% of global production, and also being affected by Covid-19 first, a global facemask shortage and a resulting so-called mass mask panic occurred at the beginning of the pandemic [6] [7]. Increased demand, coupled with panic buying, hoarding and misuse of personal protective equipment (PPE), disrupted global supply chains and put lives at risk. The demand exceeded the global production capacity, which led to significant price increases [8]. Since face masks are essential protective equipment for healthcare workers during an infectious disease outbreak, obtaining respiratory protection for clinical staff and other healthcare workers became a major global challenge. The shortages resulted in nonstandard practices in healthcare facilities, such as the usage of expired face masks, acquisition of non-certified respirators for emergency reserves as potential replacements for limited approved respirators (e. g. N95) and various alternative decontamination procedures [9] [10]. Liao et al. [11] investigated multiple commonly used disinfection schemes (e. g. heat, UV, steam, alcohol) on media with particle filtration efficiency of 95%. They concluded that heat sterilization under controlled humidity was the most promising method for the preservation of filtration properties in melt-blown fabrics and N95-grade respirators, followed by UV-radiation, that resulted in small material degradation after multiple cycles. Lastly, treatments involving steam, alcohol or bleach were characterized as leading to degradation of the filtration efficiency.

Cold atmospheric plasma (CAP) treatment, which has been investigated as a promising new approach to sterilize surfaces, particularly of heat-sensitive materials (e.g. food [12], biological material [13], contact lenses [14]) or difficult geometries (e.g. dental handpieces [15]), is a promising candidate to replace or support conventional cleaning methods and it has also been proven effective against germs of clinical interest [16]. These results suggest that CAP sterilization potentially could be an alternative decontamination method for PPE shortages, even in the clinical sector. However, it has to be considered that cold atmospheric plasma is composed of different highly reactive oxygen and nitrogen species (RONS) [17] [18], e.g. $O_3$, $NO$, $NO_2$, $NO_x$, $OH^-$, $HNO_3$, which may potentially harm material surfaces. For example, in previous research we showed three different types of reactions of various materials (stainless steel, different polymers and glass) to CAP afterglow generated by surface micro discharge (SMD) technology: no change, shift of free surface energies or oxidation [19]. In particular, PP, which is of special interest regarding face masks, showed increased polar fraction of free surface energy and an increased amount of oxygen and nitrogen species on their surface. Further research on polymers (PP) demonstrated improved wettability and wicking after plasma treatment, which afterwards decreased with storage time [20]. Knoll et al. [21] moreover determined etch rates on polymers of an atmospheric pressure plasma plume in direct or insignificant electrical interaction with the polymer surface and Bormashenko et al. [22] found the trapping of ions by polar groups of polymer surfaces – resulting in its electrical charging – to be one of the important interaction mechanisms of cold plasmas with organic surfaces by studying interactions of a cold radiofrequency plasma with low-density polyethylene substrates. Additionally, Bartis et al. researched on CAP interaction with polymers: directly after SMD plasma treatment with $N_2/O_2$ carrier gas, they found $NO_3$ groups attached to the polymer surface [23] and treatment with an atmospheric pressure plasma jet, driven with different nitrogen and oxygen admixtures, revealed material changes due to $O_3$ and $NO_x$ [24].

Triggered by the onset of the corona epidemic, first studies on alternative sterilization of face masks were conducted. A comparison of 75%-ethanol-cleaning of melt-blown and nanofiber filters demonstrated that, while both filter types maintained hydrophobicity, filtration efficacy of the melt-blown filter was significantly reduced by sterilization with ethanol, while the nanofiber filter retained its high filtration efficiency [25]. Schwan et al. [26] investigated the suitability of ozone decontamination for facepiece respirators (KN95) and concluded that a flow-through configuration provides better inactivation of viral and bacterial pathogens than the standard approach of an ozone chamber, with a 3-log reduction of *Escherichia coli* (*E. coli*) within 64 min. They also found that the treatment was non-destructive to the mask's physical structure and did not reduce filtration efficiency over time, by optical microscopy and sodium chloride aerosol test for determination of the filtration efficiency of ozone treated (30, 60, 120 min) masks. Lee et al. [27] used dielectric barrier discharge (DBD) plasma with 120 ppm ozone for inactivation of a human coronavirus (HCoV-229E) on KF94 masks and obtained a 4-log reduction by a 10s treatment. No detectable structural or functional deterioration of the material was observed by electron microscopy, particulate filtration efficiency and inhalation resistance tests after up to 5 times treatment. Furthermore it was stated, that since melt-blown filters consist of a three-dimensional network of PP fibers, which is resistant to ozone [28], any structural damage to the filter affects the inhalation resistance as well as the filtration efficiency of the mask [25] [11].

To complement the results presented, we decided on conducting in-depth analysis of long-term plasma-treated melt-blown face mask material. Since on the one hand, we found plasma impact, such as changed surface energy and attached O and N species to the polymer surface of PP in previous investigations [19] and on the other hand melt-blown-based dust masks such as KF94 or FFP3 masks contain an electrocharged filter layer, whose charge potentially could be lost with its exposure to reactive plasma components, we wanted to further study the long-term plasma effect on face masks. For this purpose, we exposed samples of FFP3 face masks to SMD atmospheric plasma for different treatment times up to 15 h and conducted material analysis by various methods, such as laser microscopy, contact angle measurements, XPS as well as filter transmittance and breathing resistance measurements. We have chosen to perform the investigations with indirect plasma treatment by SMD plasma, because in such a plasma – compared to the direct plasma treatment by jets or dielectric barrier discharges (DBD) – there is no flux of charged plasma components [29]. As a result, the plasma treatment is less effective [30] and thus associated with longer treatment times, but we also expect a lower impact on the textile. Our investigations were supplemented by a validation of cold atmospheric plasma (CAP) decontamination of FFP3 masks from *E. coli* and *Bacillus atrophaeus* (*B. atrophaeus*).

## 2 Experimental Setup, Materials and Methods

### 2.1 Experimental Setup

The SMD plasma chamber used in the experiments is enclosed in a cuboid plastic housing and has an internal volume of approx. 604 cm$^3$. The plasma in this apparatus is generated by an SMD discharge between a high voltage-driven copper electrode and a grounded aluminum grid counter electrode. A quartz glass plate, as dielectric, is located in the gap between the two electrodes. The electrode area is 181

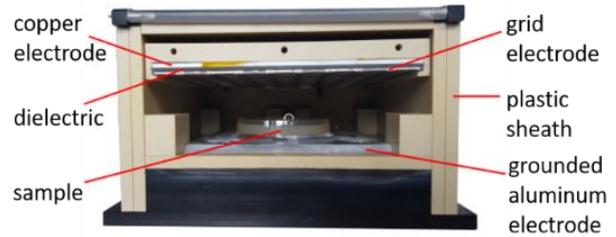

*Figure 1: Structure of the plasma chamber*

cm$^2$. The RONS, generated by the SMD discharge under atmospheric conditions, diffuse through the grounded grid electrode into the interior of the plasma chamber, where the FFP3 mask samples to be treated were located on a second grounded aluminum electrode (see Figure 1). The distance between the height-adjustable electrodes and the samples was set to 26 mm for the experiments. A plastic flap at the front of the chamber allows on the one hand the introduction and removal of samples, on the other hand it restricts an exchange with the ambient air during the plasma treatment as much as possible. The plasma apparatus was connected to a high voltage amplifier (*TREK 10/10B-HS*), which in turn was wired to a function generator (*Rohde & Schwarz HM8150*). With this configuration, a sinusoidal voltage of 12 kV$_{pp}$ was applied between the copper and aluminum grid electrodes at a frequency of 5 kHz. The voltage and current curves of the set function were observed on the linked oscilloscope (*Tektronix DPO 2012B*) (see Figure 2). At the SMD surface a power density of 0.41 W/cm$^2$ was determined via Lissajous method.[31] Regarding ozone production, the SMD plasma chamber can be run in two different modes. For each of these modes the ozone concentration (O$_3$) in the plasma chamber was determined by absorption spectroscopy. A deuterium lamp (Ocean Optics D-2000 Micropack) was used, which emits UV radiation with wavelengths between 210 and 400 nm. The radiation was directed from the source to the plasma chamber using a UV/SR-VIS fiber (Ocean Optics). Before entering the interior of the plasma device, the UV radiation was parallelized by a collimator. Ozone, generated by the plasma ignition at the electrode, diffused into the interior of the plasma chamber, through which the collimated radiation was transmitted. Since ozone absorbs electromagnetic radiation of a wavelength of 254 nm, this wavelength portion was attenuated in the beam of the deuterium lamp. After passing through the plasma chamber, the UV radiation was then captured by a second collimator and directed into a spectrometer (*Ocean Optics HR4000CG-UVNIR*). A special software (*Ocean Optics SpectraSuite*) on a laptop (*Samsung NP-Q35*) recorded the detected intensity values at 254 nm during the duration of plasma ignition. The ozone density $\rho_{O_3}$ in the plasma chamber at time t could be calculated from the recorded intensity measurements using the Beer-Lambert law:

$$\rho_{O_3}(t) = \frac{\log_{10} \frac{I_0}{I(t)}}{\varepsilon_{O_3} \cdot l}$$

Where $I_0$ is the absorbed intensity of the radiation component at 254 nm with the plasma apparatus switched off and $I(t)$ stands for the measured intensity at this wavelength with the apparatus switched on at the time of measurement t. $\varepsilon_{O_3}$ is the absorption coefficient of ozone, which is $114,7 \cdot 10^{-19}$ cm$^{-2}$, and $l$ is the distance travelled in the ozone-containing plasma. With the help of a spacer for the collimators, this distance $l$ was set to 2 cm. Ozone concentration in the plasma over time can be seen in Figure 3. Cold atmospheric plasma produced at 5 kHz and 8 kV$_{pp}$ in the SMD chamber is in so called 'ozone mode', were mostly ozone is produced in the plasma while at 5 kHz and 12 kV$_{pp}$ ozone production decreases to a minimum after a quick increase following the ignition. These observations coincide with those of Shimizu et al. [32] who concluded that the decrease of ozone concentration at high power input happens due to vibrationally excited nitrogen molecules reacting with ozone atoms and thus, generating nitric oxide. Thereby, the transition to an ozone-poisoning mode in the plasma is initiated. The plasma devolves in so-called 'nitrogen mode'. This model as well fits our experimental observations.

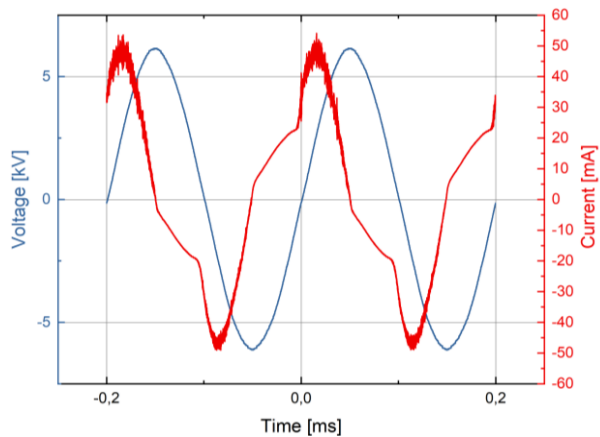

Figure 2: SMD discharge voltage and current at function settings of 5 kHz and 12 kV$_{pp}$

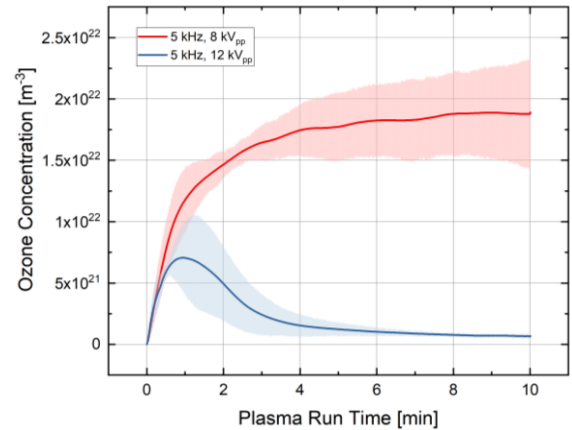

Figure 3: Ozone concentration in the SMD plasma chamber at 5 kHz and 8 kVpp / 12 kV$_{pp}$

## 2.2 FFP3 Mask Samples

As sample carriers for the inactivation experiments, pieces of the FFP3 mask measuring 1 x 1 cm$^2$ were used. Due to the different surface material structures on the inside and outside, inactivation was investigated on both sides of the mask. 100 µl of bacteria or spore solution were pipetted onto the sample carriers (see Figure 4). The bacteria used was *E. coli* (Strain L17), a well-studied gram negative, rod-shaped, non-sporulating, peritrichous flagellated bacteria that occurs in the human and animal intestines and is therefore considered a fecal indicator. [33] [34] In the human intestinal flora, *E. coli* produces vitamin K in particular. [35] Most strains of this bacterial species are nonpathogenic, but there are also numerous pathogenic strains [36], which are among the most frequent causes of human infectious diseases. The spores used were *B. atrophaeus* (Simicon® in aqua dest.), a spore-forming rod-shaped gram-positive bacterium, the majority of which is capable of active movement. *B. atrophaeus* has been established as a test germ in biological indicators for the testing of conventional sterilization processes [37], based on heat or the use of chemicals.

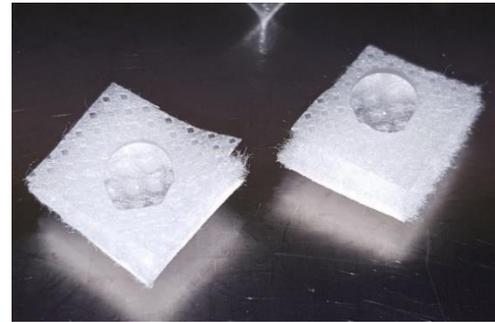

Figure 4: Inoculated FFP3 mask samples (outside)

## 2.3 Experimental Methods

### 2.3.1 Microbiological Experiments

The inactivation experiments were carried out over three days each. The day before the actual experiment, 100 µl (~10$^6$ colony forming units (cfu)) of the *B. atrophaeus* spore solution were pipetted onto the FFP3 mask samples and dried overnight. For the *E. coli* experiments, bacteria were taken from their storage in the freezer and incubated in Lysogeny Broth (LB) overnight a day prior to the experiment itself. Since living bacteria are more sensitive than spores, 100 µl (~10$^6$ cfu) *E. coli* were pipetted onto the fabric samples only on the actual day of the experiment and the samples were used immediately after drying (~ 1 h). Then, on the day of the experiment, the samples were exposed to the plasma afterglow in the SMD plasma apparatus. The chamber was run in 'nitrogen-mode' (5 kHz,12 kV$_{pp}$) since we found effective microbial inactivation in previous studies for these settings. Surviving microorganisms and cell debris were afterwards removed from the fabric in an ultrasonic bath (*Bandelin SONOREX SUPER RK 100 H*) and resolved in sterile deionized water. The resulting solution was plated out on agar plates with nutrient media (LB for *E. coli*, Reasoner's 2A agar (R2A) for *B. atrophaeus*) and incubated overnight (spread plate method) in the *Phoenix Instruments TIN-IN35* incubator. On the following day, the results of the experiments could be quantified by cfu counting and evaluated. For biological validity, the experiments were performed with three duplicates each on three different experimental days.

Since other researchers found a positive influence of high relative ambient humidity on sporicidal [38] and fungicidal [39] plasma effects, during experimentation ambient humidity and temperature were monitored. Ambient temperature in the laboratory was 24 ± 1°C at a humidity of 34 ± 1 %RH. To allow a better understanding of the experiments, the detailed steps for every experimental day (1-3) are listed below:

(1.1 a)  application of 100 µl (~$10^6$ cfu) *B. atrophaeus* to FFP3 mask samples, drying over night
(1.1 b)  cultivation of *E. coli* in LB at 37°C over night
(2.0)    application of 100 µl (~$10^6$ cfu) *E. coli* to FFP3 mask samples, drying
(2.1)    plasma treatment of inoculated FFP3 mask samples for different treatment times
(2.2)    detachment of microorganisms from the fabric samples in an ultrasonic bath (2 min) and resuspension in sterile deionised water
(2.3)    preparation of decimal dilution series from the initial solutions
(2.4)    plating out of 100 µl of the diluted suspensions onto agar with nutrient media
(2.5)    24-hour incubation at 32 °C (*B. atrophaeus*) / 37°C (*E. coli*) for cfu growth
(3.1)    cfu counting with use of a *Stuart SC6+* colony counter

*2.3.2 Surface Analysis Methods*

For sample preparation, FFP3 mask samples were long-term (15 h) plasma treated. Due to the heating of the electrode during long run times (> 30 min) and the associated necessary cooling phases (at least 30 min), this process extended over several days. These treated samples and untreated control samples were then optically examined for material damage and colour or structural changes at a macroscopic level. Additionally, untreated and long-term treated (15 h) samples of the FFP3 masks were subjected to optical analysis via imaging with a laser scanning microscope (*KEYENCE VK-9700*) and imaged with the associated software *VK Analyser* and thus could be studied on a microscopic level. The *VK-9700* operates with a violet 408 nm short waveform semiconductor laser as a light source, which, in combination with the apochromatic lenses (50x, 150x magnification), enabled ultra-high-resolution examination of the samples.

Furthermore, contact angle measurements were performed using the *OCA 20* optical contact angle measurement and contour analysis system and *SCA 20* and *SCA 21* software (*DataPhysics Instruments GmbH*). Deionized water was used as the reference liquid. The contact angles were calculated by sessile drop method. [40] Two samples each from the in- and outside of the mask material were used for control samples and for the different plasma treatment times. Contact angle measurements were performed right after termination of the long-term plasma treatment. The contact angles were measured twice on each sample - immediately after application of the droplet and one minute later.

To further determine surface changes due to plasma impact X-ray photoelectron spectroscopy was utilized. These measurements were performed with the *VersaProbe II* microscope (*Physical Electronics GmbH*) and results were imaged via *CasaXPS* software (*Casa Software Ltd* [41]). The dual-beam charge neutralizer was driven with 3 eV electrons and 10 eV argon ions. Due to higher electron flux, during the measurement, the surface potential of the samples was about -3 V. As a radiation source, a monochromatic Al-kα radiator with an energy of 1.486 eV was used. X-ray power was about 90 W and the analyzed area was 1.3 mm × 100 µm at high power mode. For the overview spectra, transmitting energy of the analyzer was 93.9 eV with a step size of 0.8 eV and an integration time of 50 ms. For the detail spectra, transmitting energy of the analyzer was adjusted to 23.5 eV and step size was 0.2 eV. Three samples each, from out- and inside of the FFP3 mask textile and of pure polypropylene (PP) and ethylene vinyl acetate (EVA), were examined.

Finally, the tests on the behaviour of plasma treated FFP3 face masks regarding permeability and breathing resistance were commissioned externally from the company *Lorenz Meßgerätebau GmbH & Co. KG*. For these investigations, 7 x 7 $cm^2$ sections of the FFP3 mask, treated with plasma for different time spans, were examined and compared to untreated control samples. Measurement of the transmittance and breathing resistance was carried out according to DIN EN 149 [42]. The sample was attached to a customized filter holder and measured on a circular surface with a diameter of 5 cm. The permeability was measured at 25 l/min for 210 s with a poly dispersive paraffin oil aerosol (20 mg/m³). Respiratory resistance was measured at three flow rates (8, 25 and 43.5 l/min.). The respective direction (inhalation or exhalation) is indicated in Figure 9 (top right) as 'IN' or 'OUT'. The volume flows were chosen so that the breathing resistances are approximately in the range of the complete mask. The factor to the normative volume flows was 3.75. The samples were measured in the filter holder, not on a Sheffield head.

## 3 Results and Discussion

### 3.1 Bacteria Experiments

The experiments showed that *E. coli* and *B. atrophaeus* could both be effectively inactivated by plasma treatment in nitrogen mode (12 kV$_{pp}$, 5 kHz). The inactivation of *B. atrophaeus* on the out- and inside of the FFP3 mask can be seen in Figure 5 (left). After 10 minutes a reduction of 1-log was achieved. After 20 minutes reduction was 2-log and after 30 minutes more than 4-log. The results for both sides of the FFP3 face mask did not differ significantly, thus, for the *E. coli* inactivation experiments only samples from the more structured outside fabric were used. *E. coli* population could be reduced by 5-log within one minute of CAP treatment (see Figure 5 (right)). After three minutes 6-log reduction was demonstrated and after five minutes a complete inactivation was achieved. Due to the logarithmic representation in Figure 5 the complete inactivation cannot be plotted at N/N0 = 0. However, it is below the detection limit at 1/N0 that can be seen in the plot. When comparing Figure 5 (left) and (right) it becomes obvious that *E. coli* reacts more sensitive to plasma afterglow species than *B. atrophaeus* with five minutes for complete inactivation compared to 30 minutes for 4-log reduction. The observation is consistent with the investigations of Klämpfl. et al. [16] and is to be expected due to spores' cellular structure with several protective cell walls.

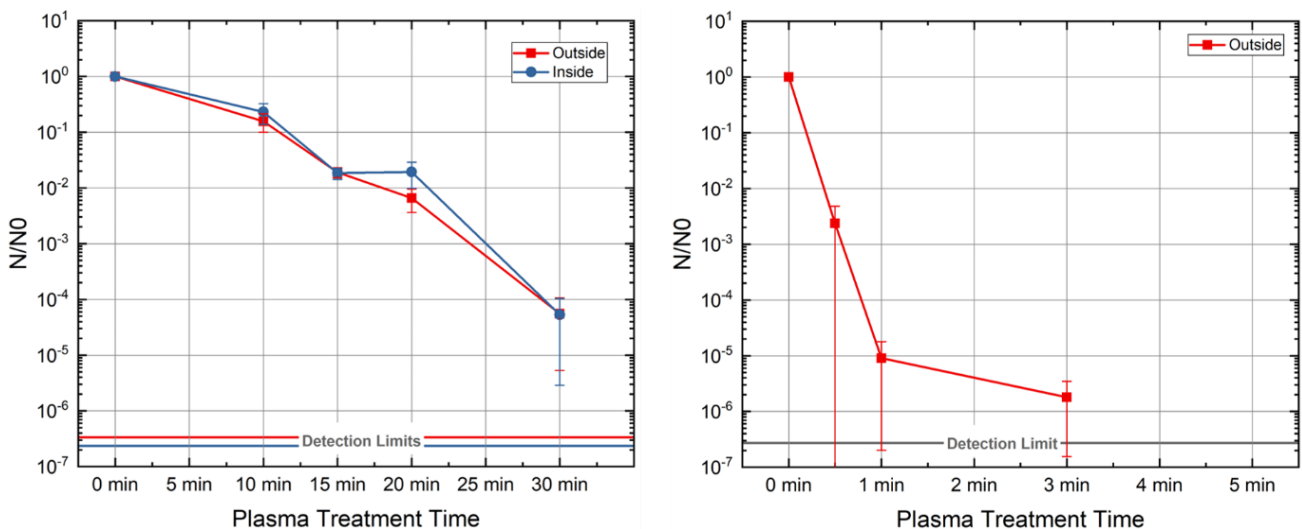

*Figure 5: Results of the inactivation experiments in nitrogen mode (12 kV$_{pp}$, 5 kHz) with standard deviations; (left) inactivation of B. atrophaeus on FFP3 face mask samples from in- and outside of the mask, (right) inactivation of E. coli on FFP3 face mask samples from the outside of the mask*

## 3.2 Surface Analysis Experiments

### 3.2.1 Optical Examination

An optical comparison of untreated and long-term treated (15 h) FF3 mask samples (see Figure 6 a)) showed no structural changes on a macroscopic level, such as material porosity or deposits. Furthermore, there are no significant deviations in colour of the fabric between the untreated (left) and long-term treated samples (right). In Figure 6 a) the 15 h plasma treated material (right) seems slightly more greyish than the untreated material (left) but the colour deviation is minimal. As the fabric of FFP3 masks is composed of EVA and PP, samples of both materials were also subjected to a long-term treatment (15 h) with plasma (see Figure 6 b) & c)). For EVA and PP no significant changes of structure or colour due to the long-term plasma treatment could be observed on the macroscopic level.

### 3.2.2 Laser Microscopic Images

Two samples each from the in- and outside of the FFP3 mask were examined. Due to the different fibre thicknesses in the structure of the surface of the outside material, magnifications (50x) were investigated in two regions (see Figure 7 area 1 & 2). In the finer-fibred area 1, no differences could be seen visually between the reference sample and the sample treated with plasma for 15 hours (see Figure 8Figure 8 a) & b)). Also, in the more coarse-fibred region 2, no significant difference between the treated and untreated sample could be detected optically (see Figure 8 c) & d)). The material does not appear to have become damaged or porous by cold plasma impact.

### 3.2.3 Contact Angle

Other research showed the enhancement of both hydrophilic and hydrophobic properties of textiles by plasma treatment in different pressure ranges and with varying carrier gases. [43] Thus, the wettability of synthetic polymers such as polyamide (PA), polyethylene (PE), polypropylene (PP), polyethylene terephthalate (PET), polytetrafluoroethylene (PTFE) etc. can be increased by oxygen, ammonia, air, nitrogen, etc. plasmas while the usage of siloxanes, perfluorocarbons, SF6, acrylates, etc. causes hydrophobic or oleophobic finishing of natural fibres

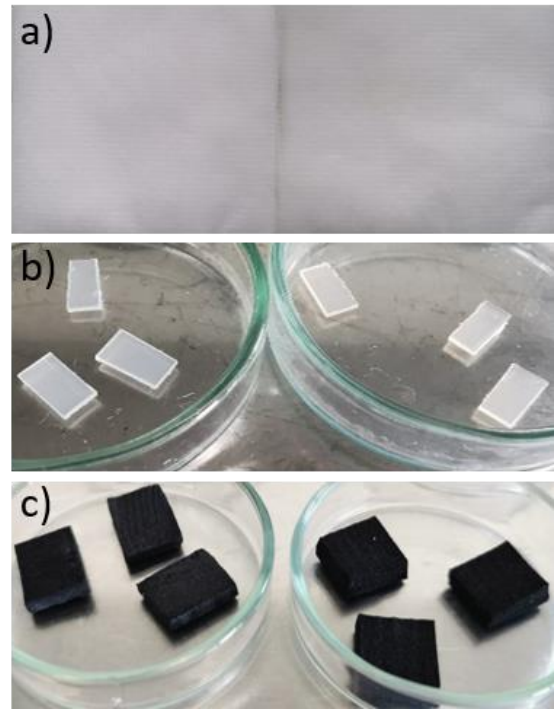

Figure 6: Optical comparison of plasma treated / untreated FFP3 mask samples and material samples; a) untreated (left) and 15 h plasma treated (right) FFP3 mask sample, b) untreated (left) and 15 h plasma treated (right) PP samples, c) untreated (left) and 15 h plasma treated (right) EVA samples

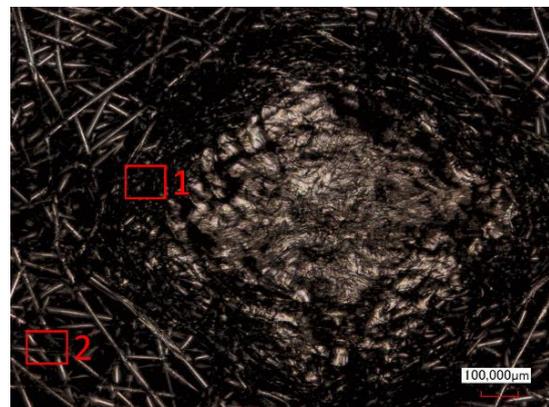

Figure 7: Laser microscope image of untreated FFP3 mask surface sample (10x magnification), locations marked in red are examples of the regions on the sample where the images in Figure 8 were taken (area 1: a) & b), area 2: c) & d))

(cotton, wool, silk, etc.). [44] For example, Samanta et al. demonstrated improved hydrophilicity in nylon and polyester by helium (He) atmospheric pressure glow discharge plasma treatment [45], Hasan et al. presented improved hydrophilicity in PP-nonwoven fibers after cold oxygen ($O_2$) plasma treatment [46], whereas K. K. Samanta et al. as well utilized a He/1,3-butadiene (BD) mixture atmospheric pressure glow plasma to reach a high degree of hydrophobic functionality in cellulosic fabric (viscose rayon) [47]. Notwithstanding these observations by other researchers, for the FFP3 mask textile – composed of PP and EVA –, despite slight variations in the results, no significant trend in the change of contact angles, and thus surface energy, with increasing cold air plasma treatment

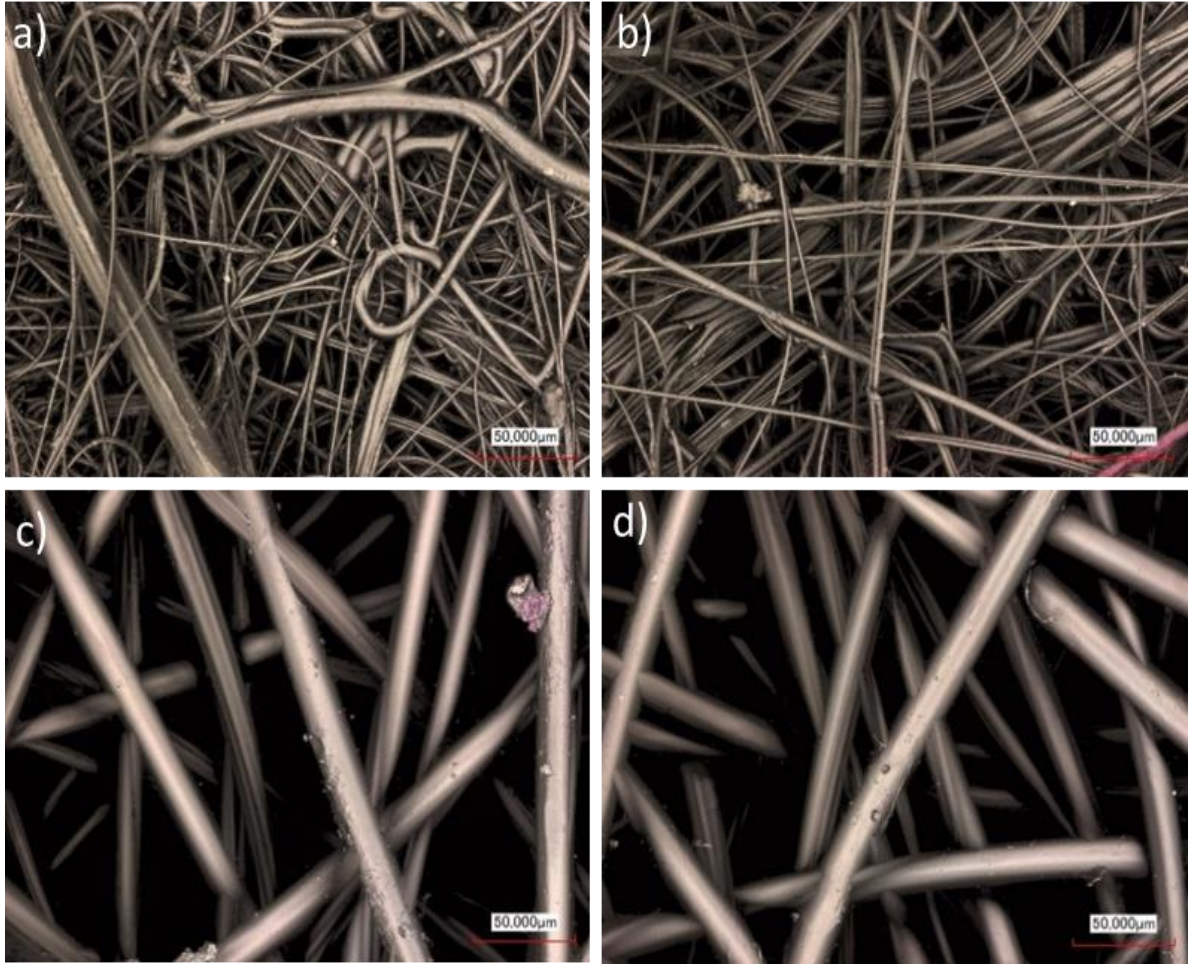

*Figure 8: Laser microscope images of plasma treated / untreated FFP3 mask surface samples; a) untreated sample (area 1, 50x magnification), b) 15 h treated sample (area 1, 50x magnification), c) untreated sample (area 2, 50x magnification), d) 15 h treated sample (area 2, 50x magnification)*

time was observed in our measurements (see Figure 9 (bottom left)). There was also no significant difference in the contact angles on the outside and inside of the FFP3 mask material. The measurement results were averaged in each case and standard deviations were calculated.

### 3.2.4 XPS

As ca be seen in Figure 9 (bottom right) XPS measurements determined a surface composition of 94.12% carbon (C), 5.01% oxygen (O), 0.47% calcium (Ca) and 0.41% silicon (Si) for the untreated control sample (red curve). After 15h of afterglow plasma treatment the surface composition of the fabric samples has not changed, as to be seen in Figure 9 (bottom right, green curve). In addition to the FFP3 mask samples, one sample each of the pure components (EVA & PP) of the mask material were analyzed. Considering the sum formulas of EVA $((C_2H_4)_n(C_4H_6O_2)_m)$ and PP $((C_3H_6)_n)$ the presence of carbon, hydrogen and oxygen (for EVA) was to be expected. However, hydrogen is not found in the spectrum and oxygen was found in the PP spectra of both, reference and plasma-treated sample. The absence of hydrogen results from the measurement method. XPS measurements only work from the effective cross-section of lithium, hydrogen has a much too small effective cross-section and can therefore not be measured. The presence of oxygen on the PP reference sample can be explained by impurities, i.e. hydrocarbons, on the material sample. Furthermore, a comparison of PP spectra revealed a decrease of C and increase of O due to plasma treatment, where there is no CO fraction in the C peak, so no clear evidence of oxidation of C. Detected calcium and silicon are presumably impurities in the plastics industry. From each FFP3 mask sample 3 specimens were measured, each with an overview spectrum and detail spectra of the C, O and Si peaks. For the detail spectra, a variation in the half-width of the peaks was observed between the 3 samples. These can probably be explained by the trapping of argon ions during the XPS measurement and a resulting charging of the mask surface, which could not be fully compensated by the ion and electron neutralizers due to the 3D structure of the mask surface. For both pure EVA and PP, a nitrogen (N) peak was measured at 402 eV in the long-term plasma-treated (15h) samples, which may have resulted from the accumulation of plasma components on the material surface. This peak was not present in the untreated EVA and PP samples. No N peak could be measured in the FFP3 mask samples either. This can possibly be explained by a surface effect: because the mask material has a very large surface area due to its fiber structure, an occurrence of the same number of N on the surface could possibly be so widely distributed on these samples that they cannot be significantly detected in the measurement. Finally, a zinc (Zn) peak was measured in treated and untreated EVA samples and fluorine (F) was measured only in the plasma-treated EVA sample. These two elements probably indicate impurities: fluorine, for example, occurs in sealing materials of vacuum systems.

*3.2.5 Permeability and Resistance Measurements*

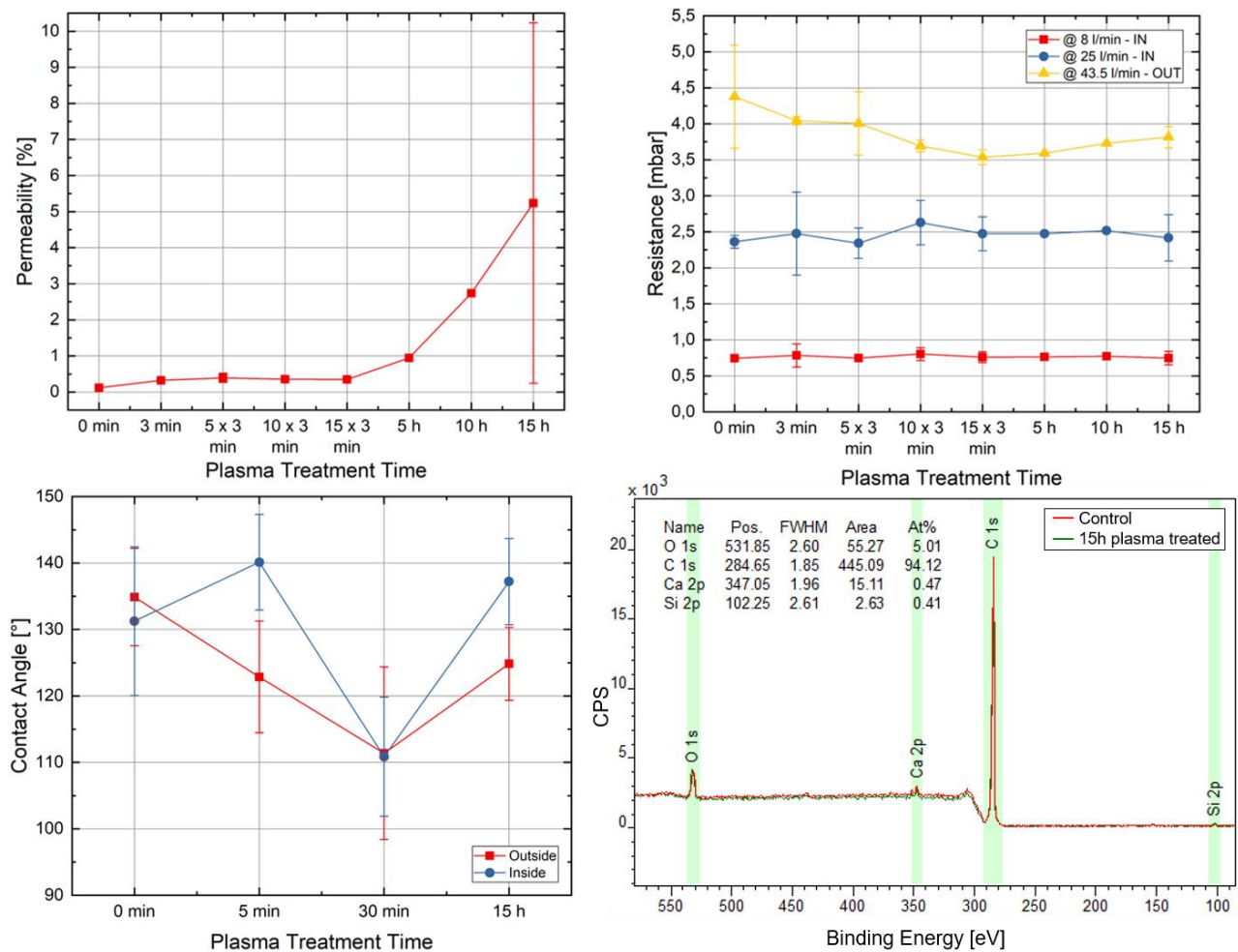

*Figure 9: Results ((top & bottom left) averaged, with standard deviation errors) of experiments for determination of fabric properties of untreated / plasma treated FFP3 face mask samples; (top left) Permeability, (top right) Respiratory resistance, (bottom left) contact angles of deionised water on the material surface, (bottom right) XPS measurement spectra normalized at X = 1123.08 of FFP3 face mask outside*

The examined treatment times were chosen to investigate material reaction to long-term treatments (5, 10 and 15 h) and treatment times that would be realistic in everyday use by private individuals (3, 5 x 3, 10 x 3 and 15 x 3 min). For control and each treatment time, besides 5h and 10 h plasma treatment, two samples each were measured and the results were averaged. Results are shown in Figure 9 (top left/right). As can be seen in Figure 9 (top left), a clear increase in permeability of the mask material was detected with increasing plasma treatment time. For the 3, 5 x 3, 10 x 3 and 15 x 3 min treatments, the permeability was slightly increased, but still below 1%. For the 5h treatment permeability was 0.95%. For the long-term treatments (10 h & 15 h), the transmission rate increased to 2.74% and 5.24% min respectively. In contrast, a clear change in the measurement results with increasing treatment time was not observed for the breathing resistance at any of the flow rates (8 l/min - IN, 25 l/min - IN, 43.5 l/min - OUT) (see Figure 9 (top right)). Thus, our results show that the plasma treatment resulted in a significant increase in the mask material's permeability, while no significant change in breathing resistance was observed. This discrepancy can be explained by the fact that the breathing resistance is due to purely mechanical effects, whereas the permeability of melt-blown-based filter media, is based on a combination of mechanical and electrostatic properties of the face mask material. The results seem to indicate that the plasma treatment reduces the charging of the material with increasing treatment duration while no mechanical damage to the material occurs, so that the permeability increases accordingly, but the breathing resistance remains unaffected. The electrostatic filtering layer allows to capture the finest aerosols and to retain viruses that are not captured by the pure fibre density of the fabric. The observed decrease in the charge of the filter layer could be explained by the trapping of charged plasma components and corresponding neutralization. None of the other material measurements performed contradicts this hypothesis and it goes hand in hand with considerations that state that any structural damage to the filter affects the inhalation resistance as well as the filtration efficiency of the mask [11] [25].

## 4 Conclusion

Our studies have shown that germ inactivation on FFP3 masks is possible by cold atmospheric SMD plasma. An efficient reduction of the model organism *E.coli* by more than six orders of magnitude could be achieved within 4 min of plasma treatment. Inactivation of *B. atrophaeus* endospores, a typical biological indicator for industrial detection of sterilization, on the mask material took significantly longer: a reduction of over four powers of ten was achieved by a 30 min plasma treatment.

By the material analysis methods optical analysis, laser microscopy, contact angle measurement and XPS, no significant influence of long-term plasma treatment (15 h) on the FFP3 facemask material could be detected. Optically, only minimal colour deviations of the mask material due to plasma treatment could be observed. No colour differences or damage to the surfaces of the individual material samples of EVA and PP could be determined. Laser microscopic analysis of the mask samples also did not reveal any changes in the mask material at a microscopic level. Furthermore, despite slight variations in the results, no significant trend in the change of contact angles, and thus surface energy, with increasing cold air plasma treatment time was observed in our measurements. The XPS measurements as well did not reveal any clear material changes in the melt-blown fiber material, such as oxidation of certain surface molecules or clear accumulation of foreign substances by the plasma. However, changes in the individual spectra of PP and EVA were detected, such as the decrease of C and an increase of O in the plasma long-term treated sample of pure PP or the accumulation of N on the plasma treated PP and EVA samples. The accumulation of N could be explained by plasma components, while the change in C and O peaks are difficult to explain due to the lack of oxidation components in the peaks. The most distinct influence of the 15-hour long-term plasma treatment on the FFP3 mask material was found in the filter performance measurements: the plasma treatment resulted in a significant increase in the mask material's permeability, while no significant change in breathing resistance was observed. From these results it can be deduced that the material itself is not damaged by the plasma, but that the electrostatic filtering effect of the FFP3 mask is weakened over time by the plasma treatment.

Our results suggest that the reuse of plasma sterilized FFP3 masks is possible, but to a limited extent. The cut-off values for the transmittance of FFP3 masks are 99% [42], while typical transmittances for e.g. factory-new FFP3 masks are mostly below 0.1%. This could also be observed in our control sample (see Figure 9 (top left)). A small increase in the permeability, although still well within the tolerance range, can already be seen after a three-minute plasma treatment but the threshold value of 1% permeability is only reached after a 5-hour plasma treatment of the mask material. Thus, according to our measurements, it should be safe to decontaminate FFP3 masks by less than 5 hours CAP afterglow treatment without exceeding the permeability limit in the case of PPE shortages, which can be experienced during pandemics.